\documentclass{an_art}
\usepackage{epsfig}
\usepackage{graphicx}
\usepackage{amssymb}
\def\rfnote{\small\parindent=.0cm\hangindent=20pt\hangafter=1\indent\par}

\begin{document}
\begin{titlepage}
\setcounter{page}{347}

\headnote{Astron.~Nachr.~319 (1998) 6, 347 -- 368}
\makeheadline
\title{The ROSAT Bright Survey:\\[0.5ex]
I. Identification of an AGN sample with hard ROSAT X-ray spectra
\thanks{Based on observations at the European Southern Observatory
        La Silla (Chile) with the $\rm 2.2m$ telescope of the 
	Max-Planck-Society}}
\author{{\sc J.~-U.~Fischer, G.~Hasinger, A.~D.~Schwope, H.~Brunner},
Potsdam, Germany \\
\medskip
{\small Astrophysikalisches Institut Potsdam} \\
\bigskip
 {\sc T.~Boller, J.~Tr\"umper, W.~Voges}, Garching, Germany \\
 \medskip
 {\small Max-Planck-Institut f\"ur extraterrestrische Physik}\\
\bigskip
 {\sc S. Neizvestny}, Nizhnij Arkhyz, Russia \\
 \medskip
 {\small Special Astrophysical Observatory}
}
\date{Received 1998 September 24; accepted 1998 October 29} 
\maketitle
%
\summary
The ROSAT Bright Survey (RBS) aims to completely optically identify the more
than 2000 brightest sources detected in the ROSAT all-sky survey at galactic
latitudes $\rm |b|>30\degr$ (excluding LMC, SMC, Virgo cluster).
This paper presents a subsample of 66 bright point-like ROSAT survey 
sources with almost hard PSPC spectra, the hardness ratio HR1 is $> 0.5$ for
most of the sources. The subsample 
could be nearly completely identified 
by low-resolution optical spectroscopy with the following
breakdown into object classes: 31
Seyfert galaxies, 22 BL Lac candidates,
5 clusters of galaxies, 1 cataclysmic variable, and 5 bright stars.
Only one object remained unidentified and one X-ray source was a spurious 
detection.
The redshift distribution peaks around 0.06 for the Seyferts and around 0.13
for the BL Lac candidates. 
Observations with medium spectral resolution were obtained for most 
of the new Seyfert galaxies. 
A large fraction (20 objects) are type 1 Seyfert galaxies, 
the other fraction includes Seyfert galaxies of type 1.5 -- 1.8 (5 objects), 
two LINERs, and 4 possible narrow-line Seyfert 1 galaxies (NLS1). 
About one third of the  new Seyfert's have nearby companion galaxies 
displaying either emission or absorption lines at the same redshift. 
Among them are a couple of systems showing 
direct morphological evidence for interaction. 
The large fraction of interacting galaxies among our sample 
suggests a scenario where interaction is the main trigger of AGN activity.
END
\keyw
AGN -- interacting galaxies -- surveys: X-ray -- surveys: AGN
END
\end{titlepage}
\renewcommand{\thempfootnote}{\alph{mpfootnote}}
\renewcommand{\thefootnote}{\alph{footnote}}

\section{Introduction}
\label{sect_intro}
Active galactic nuclei (AGNs) of different types are the major contributors
to the extragalactic X-ray background (XRB). 
According to recent models by e.g.~Comastri et al.~(1995) the spectrum 
of the XRB can be reproduced over the broad energy range 5 -- 100\,keV by 
AGN populations with different amounts of absorption, $N_{\rm H}$ up to 
$10^{25}$\,cm$^{-2}$, which undergo strong cosmological evolution. In addition,
Compton reflection which is likely associated with the
accretion disk, plays an important role in shaping the XRB-spectrum 
(e.g.~Pounds et al.~1990). 

Major uncertainties in synthetic XRB-spectra
are introduced by uncertain number-flux relations, luminosity functions 
and their cosmological evolution. Major observational impact, therefore,
comes from the definition of statistically significant samples of AGNs
using X-ray data. The first such sample (Piccinotti et al.~1982), defined
using HEAO-1 data, comprised only 34 AGN (energy band 2 -- 10\,keV). The soft
X-ray band has been opened by the Einstein observatory and the optical 
identifications of the Extended Medium Sensitivity Survey sources 
(EMSS, Gioia et 
al.~1990, Stocke et al.~1991, Maccacaro et al.~1994), 
which revealed about 450 AGNs in the 
0.3 -- 3.5\,keV passband. The launch of ROSAT finally, 
the all-sky survey (RASS) 
performed in 1990/1991 and the ongoing guest investigator program 
using both detectors onboard, the PSPC and the HRI, allowed to identify 
large samples of AGNs from wide-angle to shallow surveys using the same
well-calibrated instrument in the passband 0.1 -- 2.4\,keV. 
The deepest survey has been performed in
the {\it Lockman Hole} 
reaching a flux limit of 
$5.5 \times 10^{-15}$\,erg cm$^{-2}$ s$^{-1}$ 
in a $0.3\deg^2$ field (Hasinger et al.~1998, Schmidt et al.~1998). 
The surveys performed by Mason et al.~(1998; RIXOS, see also
Puchnarewicz et al.~1996, 1997) and Appenzeller et al.~(1998; see also
Zickgraf et al.~1997) reached flux limits of 
$3 \times 10^{-14}$\,erg cm$^{-2}$ s$^{-1}$ and
$2-10 \times 10^{-13}$\,erg cm$^{-2}$ s$^{-1}$ 
in $14\deg^2$ and $685\deg^2$ fields, respectively. A first all-sky sample 
drawn from the RASS 
with soft PSPC-spectra and countrates $> 0.5$\,s$^{-1}$ have very
recently been published by Thomas et al.~(1998), another large-area sample 
based on the RASS has been published by Bade et al.~(1995, 1998) 
with a survey area of 8480 deg$^2$. The latter authors 
correlated the RASS with sources identified on Schmidt plates 
exposed in the framework of the Hamburg Quasar Survey HQS.

Here we report on the identification of another sample of point-like 
X-ray sources
drawn from the RASS with relatively hard X-ray spectra (in ROSAT
terms). This work was started as a  pilot program for the larger
project of identifying all bright, high-galactic latitude X-ray sources 
found in the RASS, the so-called ROSAT Bright Survey (RBS), 
which aims to the complete identification of all bright X-ray sources
irrespective their X-ray colour and their X-ray extent. 

This paper is organized in the following way: In chapter 2 
we describe the X-ray properties of our sample, the 
selection strategy for sources to be identified spectroscopically 
and give an identification summary for previously
known sources. Chapter 3 gives details of our optical 
observations and the analysis. Our results are summarized 
in chapter 4 where positions and identifications  
of the optical counterparts of the X-ray sources are listed. 
Finding charts and optical spectra of the new objects are 
presented in the appendix.
Preliminary results of this work have been described in Hasinger et al.~(1997).

\begin{figure}[thb]
\resizebox{115mm}{!}
{\includegraphics[bbllx=56pt,bblly=74pt,bburx=533pt,bbury=572pt,angle=-90,clip=]
{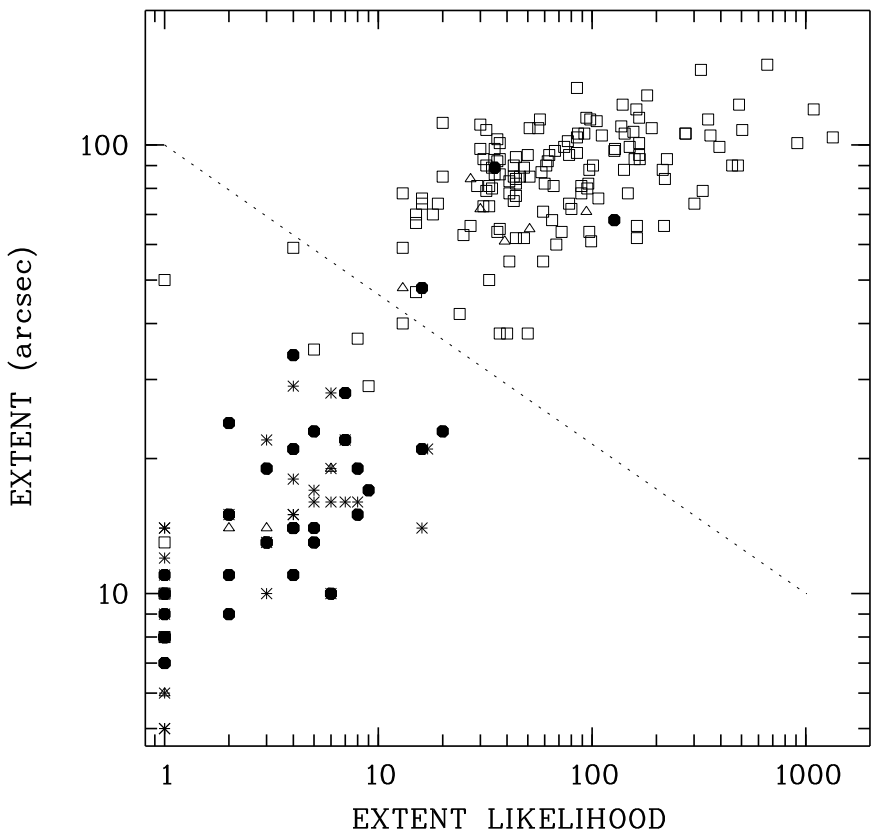}}
\hfill
\parbox[t]{55mm}{
\caption{
Discrimination between point-like and extended X-ray sources for the
catalogued X-ray sources with hardness ratio HR1 $> 0.5$ and count rate 
$> 0.2$\,s$^{-1}$. 
Legend: 
$\bullet$ AGN, $\triangle$ galaxy, $\square$ galaxy cluster, $\ast$ star.
The source extent and its likelihood were calculated
in the ROSAT standard analysis system (RASS) using a maximum likelihood source
detection and para\-meter estimation algorithm. Extended sources appear
at top right (the AGNs in that region of the diagram are superposed
on clusters). Only sources in the lower left of the diagram were
considered for identification work.
}
\label{f:ext}}
\end{figure}

\section{Selection of X-ray sources}
\label{sect_sel}
The work described here is embedded in a larger project, the ROSAT Bright
Survey (RBS), which aims to establish a complete, X-ray selected, flux-limited,
all-sky sample of AGN's. This aim is achieved by 
optical identification of all point-like X-ray sources at
high galactic latitudes and above a certain X-ray countrate during the 
ROSAT all-sky survey (RASS).
Inspection of catalogued ROSAT X-ray sources revealed, 
that the fraction of AGN's
in a sample of X-ray sources 
becomes higher with harder X-ray color.
In the first step of the whole project with the results presented
here we tried to maximise the likely AGN content of the subsample
by applying a hardness ratio criterion for HR1. 
The hardness ratio HR1 is defined as
\mbox{(H+S)/(H--S)}, with the ROSAT-PSPC count rates H in the hard band
($0.5\ldots2.0$ keV) and S in the soft band ($0.1\ldots0.4$ keV).
Our original subsample was based on the SASS 1 processed RASS and comprised
all sources at high galactic latitude, $|b^{II}|>30\degr$, brighter than
$>0.2$ PSPC s$^{-1}$ with hardness ratio HR1 $>0.5$. 
We excluded sources within circles
of radius $5\degr$ centred on ($\alpha, \delta$) = ($16\fd44, -73\fd27$), 
($83\fd80, -68\fd00$), and ($188\fd30, 10\fd70$), respectively, i.e.%
sources within directions to the
Magellanic clouds and the Virgo cluster.
\renewcommand{\baselinestretch}{0.85}\noindent\small
\begin{table}[ht]
\begin{center}
\caption{\label{t:xdat} List of X-ray sources forming the ROSAT Hard Survey RHS.
X-ray count rates and hardness ratios are derived 
from the All-Sky Survey Bright Source Catalogue (1RXS, Voges et al.~1996).
Column D(etector) indicates the detector used for position determination 
(P$=$PSPC, H$=$HRI).
}
\bigskip

\small
\begin{tabular}{|c|c||l|ccc|ccc|c|c||r@{ $\pm$ }l||r@{ $\pm$ }l|}
\hline
&&& \multicolumn{8}{c||}{X-ray Position}&\multicolumn{2}{c||}{Count Rate}&\multicolumn{2}{c|}{Hardness}\\
RHS & RBS & 1RXS\,J\dots & \multicolumn{3}{c|}{RA(2000)} & \multicolumn{3}{c|}{DE(2000)}& D & $1\sigma$ &\multicolumn{2}{c||}{CR}&\multicolumn{2}{c|}{ratio}  \\
No.~&No.~&&$h$&$m$&$s$&$^\circ$&$'$&$''$&&$''$&\multicolumn{2}{c||}{$cts/sec$}&\multicolumn{2}{c|}{HR1}\\[0.5ex] \hline
 01 & 0042 & $001827.8+294730$ & 00 & 18 & 27.8 &$ +29 $& 47 & 30 & P & 07 & 0.46 & 0.04 & 0.36 & 0.07 \\
 02 & 0045 & $002020.7+103451$ & 00 & 20 & 20.7 &$ +10 $& 34 & 51 & P & 13 & 0.31 & 0.04 & 0.74 & 0.08 \\
 03 & 0078 & $003422.2-790525$ & 00 & 34 & 22.2 &$ -79 $& 05 & 26 & P & 10 & 0.51 & 0.05 & 0.57 & 0.09 \\
 04 & 0082 & $003514.9+151513$ & 00 & 35 & 14.9 &$ +15 $& 15 & 13 & P & 08 & 0.24 & 0.02 & 0.35 & 0.11 \\
 05 & 0094 & $004053.2-074201$ & 00 & 40 & 53.2 &$ -07 $& 42 & 02 & P & 08 & 0.42 & 0.04 & 0.34 & 0.08 \\
 06 & 0104 & $004351.2+242422$ & 00 & 43 & 53.3 &$ +24 $& 24 & 22 & H & 06 & 0.50 & 0.05 & 0.75 & 0.06 \\
 07 & 0140 & $005817.1+172300$ & 00 & 58 & 17.1 &$ +17 $& 23 & 00 & P & 08 & 0.33 & 0.03 & 0.74 & 0.07 \\
 08 & 0157 & $010908.3+181601$ & 01 & 09 & 08.0 &$ +18 $& 16 & 10 & H & 05 & 0.38 & 0.04 & 0.65 & 0.07 \\
 09 & 0161 & $011050.0-125455$ & 01 & 10 & 50.0 &$ -12 $& 54 & 56 & P & 07 & 0.75 & 0.07 & 0.34 & 0.08 \\
 10 &      &        $$         & 01 & 25 & 58.1 &$ +15 $& 18 & 14 & P & 10 & 0.21 & 0.02 & 0.58 & 0.12 \\
 11 & 0258 & $015600.2+150206$ & 01 & 56 & 00.2 &$ +15 $& 02 & 06 & P & 09 & 0.31 & 0.03 & 0.69 & 0.08 \\
 12 & 0279 & $020640.5-714822$ & 02 & 06 & 40.0 &$ -71 $& 48 & 16 & H & 05 & 0.23 & 0.06 & 0.89 & 0.20 \\
 13 & 0281 & $020702.5+293035$ & 02 & 07 & 02.5 &$ +29 $& 30 & 35 & P & 10 & 0.20 & 0.03 & 0.81 & 0.10 \\
 14 & 0298 & $021632.3+231448$ & 02 & 16 & 32.3 &$ +23 $& 14 & 48 & P & 10 & 0.38 & 0.04 & 0.82 & 0.07 \\
 15 & 0345 & $024215.2+053037$ & 02 & 42 & 15.2 &$ +05 $& 30 & 37 & P & 08 & 0.49 & 0.05 & 0.56 & 0.09 \\
 16 &      & $024524.3+143518$ & 02 & 45 & 24.3 &$ +14 $& 35 & 18 & P & 10 & 0.18 & 0.03 & 0.61 & 0.14 \\
 17 & 0379 & $030007.6+163023$ & 03 & 00 & 07.6 &$ +16 $& 30 & 24 & P & 07 & 0.33 & 0.03 & 0.91 & 0.04 \\
 18 & 0381 & $030136.9+015517$ & 03 & 01 & 38.1 &$ +01 $& 55 & 20 & H & 06 & 0.21 & 0.03 & 0.90 & 0.07 \\
 19 & 0384 & $030330.0+055425$ & 03 & 03 & 30.0 &$ +05 $& 54 & 25 & P & 10 & 0.27 & 0.03 & 0.87 & 0.07 \\
 20 & 0400 & $031422.7+062001$ & 03 & 14 & 23.7 &$ +06 $& 19 & 58 & H & 05 & 0.53 & 0.04 & 0.96 & 0.02 \\
 21 &      & $031611.4+090445$ & 03 & 16 & 12.6 &$ +09 $& 04 & 46 & H & 05 & 0.14 & 0.02 & 0.87 & 0.08 \\
 22 & 0425 & $032738.6-580937$ & 03 & 27 & 38.6 &$ -58 $& 09 & 38 & P & 08 & 0.22 & 0.03 & 0.78 & 0.10 \\
 23 & 0444 & $033424.5-151325$ & 03 & 34 & 24.5 &$ -15 $& 13 & 26 & P & 09 & 0.29 & 0.03 & 0.49 & 0.07 \\
 24 & 0487 & $035257.7-683120$ & 03 & 52 & 57.7 &$ -68 $& 31 & 20 & P & 08 & 0.58 & 0.08 & 0.92 & 0.06 \\
 25 &      & $044154.5-082639$ & 04 & 41 & 54.5 &$ -08 $& 26 & 39 & P & 08 & 0.20 & 0.02 & 0.80 & 0.07 \\
 26 & 0665 & $053527.5-432247$ & 05 & 35 & 27.5 &$ -43 $& 22 & 47 & P & 08 & 0.21 & 0.03 & 0.42 & 0.12 \\
 27 & 0679 & $054357.3-553206$ & 05 & 43 & 57.3 &$ -55 $& 32 & 06 & P & 07 & 0.69 & 0.03 & 0.67 & 0.03 \\
 28 & 0702 & $082713.6+412830$ & 08 & 27 & 13.6 &$ +41 $& 28 & 30 & P & 07 & 0.23 & 0.02 & 0.53 & 0.09 \\
 29 & 0712 & $083811.0+245336$ & 08 & 38 & 11.0 &$ +24 $& 53 & 36 & P & 12 & 0.21 & 0.03 & 0.74 & 0.10 \\
 30 & 0797 & $094713.2+762317$ & 09 & 47 & 13.2 &$ +76 $& 23 & 18 & P & 08 & 0.27 & 0.03 & 0.78 & 0.06 \\
 31 & 0921 & $105607.0+025215$ & 10 & 56 & 07.0 &$ +02 $& 52 & 15 & P & 09 & 0.59 & 0.06 & 0.76 & 0.06 \\  
 32 & 0980 & $112407.9+061256$ & 11 & 24 & 07.8 &$ +06 $& 12 & 56 & P & 08 & 0.42 & 0.03 & 0.54 & 0.06 \\
 33 & 1028 & $114516.1+794054$ & 11 & 45 & 15.2 &$ +79 $& 40 & 53 & H & 05 & 0.37 & 0.03 & 0.79 & 0.04 \\
 34 & 1068 & $120711.1-174551$ & 12 & 07 & 11.5 &$ -17 $& 46 & 02 & H & 05 & 0.22 & 0.05 & 1.00 & 0.26 \\
 35 & 1080 & $121321.6-261802$ & 12 & 13 & 23.4 &$ -26 $& 18 & 08 & H & 05 & 0.29 & 0.04 & 0.82 & 0.10 \\
 36 & 1161 & $124843.3+065426$ & 12 & 48 & 43.1 &$ +06 $& 54 & 25 & H & 08 & 0.41 & 0.05 & 0.44 & 0.11 \\
 37 & 1279 & $133113.6-252406$ & 13 & 31 & 13.6 &$ -25 $& 24 & 06 & P & 07 & 0.67 & 0.05 & 0.63 & 0.06 \\
 38 & 1329 & $135420.2+325547$ & 13 & 54 & 20.2 &$ +32 $& 55 & 48 & P & 07 & 0.83 & 0.05 & 0.36 & 0.05 \\
 39 & 1372 & $141922.5-263842$ & 14 & 19 & 22.2 &$ -26 $& 38 & 35 & H & 05 & 1.22 & 0.09 & 0.46 & 0.06 \\
 40 & 1411 & $143703.5+234236$ & 14 & 37 & 03.5 &$ +23 $& 42 & 36 & P & 07 & 0.25 & 0.03 & 0.57 & 0.08 \\
 41 & 1424 & $144505.9-032613$ & 14 & 45 & 05.9 &$ -03 $& 26 & 14 & P & 08 & 0.25 & 0.03 & 0.68 & 0.10 \\
 42 & 1429 & $144725.6+082733$ & 14 & 47 & 25.6 &$ +08 $& 27 & 33 & P & 10 & 0.28 & 0.03 & 0.50 & 0.10 \\
 43 & 1431 & $144739.2-083331$ & 14 & 47 & 38.6 &$ -08 $& 33 & 23 & H & 05 & 0.25 & 0.03 & 0.73 & 0.09 \\
 44 & 1434 & $144932.3+274630$ & 14 & 49 & 32.9 &$ +27 $& 46 & 22 & H & 05 & 0.42 & 0.03 & 0.49 & 0.06 \\
 45 &      & $150636.4-054011$ & 15 & 06 & 36.4 &$ -05 $& 40 & 11 & P & 10 & 0.14 & 0.02 & 0.51 & 0.13 \\
 46 & 1467 & $150842.2+270910$ & 15 & 08 & 42.5 &$ +27 $& 09 & 09 & H & 05 & 0.33 & 0.03 & 0.54 & 0.08 \\
 47 & 1478 & $151618.7-152347$ & 15 & 16 & 18.7 &$ -15 $& 23 & 47 & P & 08 & 0.40 & 0.04 & 0.94 & 0.03 \\
 48 & 1497 & $152346.8-004434$ & 15 & 23 & 46.0 &$ -00 $& 44 & 27 & H & 05 & 0.32 & 0.04 & 0.89 & 0.06 \\
 49 & 1499 & $152451.4-100606$ & 15 & 24 & 51.4 &$ -10 $& 06 & 06 & P & 08 & 0.31 & 0.03 & 0.79 & 0.06 \\
 50 & 1505 & $153049.4+202601$ & 15 & 30 & 49.4 &$ +20 $& 26 & 02 & P & 09 & 0.26 & 0.03 & 0.49 & 0.11 \\
 51 & 1507 & $153140.9+201927$ & 15 & 31 & 40.9 &$ +20 $& 19 & 27 & P & 09 & 0.23 & 0.03 & 0.60 & 0.10 \\
 52 & 1509 & $153253.7+302103$ & 15 & 32 & 53.7 &$ +30 $& 21 & 04 & P & 08 & 0.24 & 0.03 & 0.75 & 0.09 \\
 53 & 1524 & $154016.4+815504$ & 15 & 40 & 16.4 &$ +81 $& 55 & 04 & P & 07 & 0.49 & 0.03 & 0.38 & 0.05 \\
 54 & 1620 & $170245.5+725330$ & 17 & 02 & 45.5 &$ +72 $& 53 & 30 & P & 07 & 0.41 & 0.02 & 0.70 & 0.03 \\
 55 & 1634 & $171706.8+293117$ & 17 & 17 & 07.1 &$ +29 $& 31 & 21 & H & 06 & 0.28 & 0.02 & 0.72 & 0.06 \\
 56 & 1688 & $203927.2-301844$ & 20 & 39 & 27.0 &$ -30 $& 18 & 50 & H & 05 & 0.43 & 0.04 & 0.65 & 0.07 \\
 57 & 1787 & $215015.6-141047$ & 21 & 50 & 15.6 &$ -14 $& 10 & 48 & P & 10 & 0.58 & 0.05 & 0.70 & 0.07 \\
 59 & 1882 & $224017.7+080316$ & 22 & 40 & 17.3 &$ +08 $& 03 & 14 & H & 05 & 0.48 & 0.05 & 0.70 & 0.07 \\
 60 & 1888 & $224341.9-123106$ & 22 & 43 & 41.9 &$ -12 $& 31 & 06 & P & 11 & 0.30 & 0.07 & 0.49 & 0.20 \\
 61 & 2005 & $232554.6+215310$ & 23 & 25 & 54.6 &$ +21 $& 53 & 10 & P & 07 & 0.56 & 0.04 & 0.63 & 0.05 \\
 62 & 2036 & $234106.5+093805$ & 23 & 41 & 06.5 &$ +09 $& 38 & 05 & P & 08 & 0.21 & 0.02 & 0.87 & 0.07 \\
 63 & 2040 & $234312.3-363807$ & 23 & 43 & 13.5 &$ -36 $& 37 & 50 & H & 05 & 0.24 & 0.05 &-0.58 & 0.17 \\
 64 & 2045 & $234728.8+242743$ & 23 & 47 & 28.8 &$ +24 $& 27 & 44 & P & 07 & 0.26 & 0.02 & 0.75 & 0.05 \\
 65 & 2051 & $235018.0-055928$ & 23 & 50 & 18.0 &$ -05 $& 59 & 28 & P & 08 & 0.32 & 0.03 & 0.53 & 0.09 \\
 66 & 2061 & $235547.6+253044$ & 23 & 55 & 48.6 &$ +25 $& 30 & 32 & H & 05 & 0.26 & 0.03 & 0.61 & 0.10
\\[0.5ex]
\hline
\end{tabular}
\normalsize

\end{center}
\end{table}
\renewcommand{\baselinestretch}{1}\normalsize\clearpage
\noindent

The distribution of objects on the celestial sphere is shown in Fig.~1
in Hasinger et al.~(1997). The whole survey area is 20\,391\,deg$^2$.
After elimination of 37 sources in the Virgo, LMC, SMC regions we were
left with 351 sources. Their positions
were correlated with the major optical, radio and
infrared catalogues using SIMBAD and the NED. A large fraction of
76\% (268 sources) could be identified this way with the following breakdown into object
classes: 140 clusters of 
galaxies, 
46 AGNs, 
23 unspecified galaxies,
11 nearby normal galaxies,
34 active stars (including coronally active stars, CVs and XRBs),
and 14 SNRs (see Hasinger et al.~1997).
Only 83 objects remained unidentified or were catalogued as galaxies 
(active or inactive) but without measured redshifts.
Up to this point all bright, hard RASS X-ray sources were considered.
Among these 83 sources a number of clusters were 
expected. We excluded these by application of an X-ray extent 
criterion (Fig.~\ref{f:ext}). The SASS reveals two quantities 
addressing the extent of an X-ray source, the measured extent and the 
extent likelihood (see Voges et al.~1996). By plotting these quantities 
for the identified sources a clear distinction between the extended
X-ray sources (galaxies, clusters) in the upper right and the 
point-like X-ray sources in the lower left part of the diagram 
could be established. The dashed line in the figure defines our
dividing line, only sources below this line were considered for 
optical identification, 15 cluster candidates above the line are
excluded. Two further sources turned out to be X-ray ghosts 
by inspection of the X-ray images derived from the RASS.
Finally, we were left with 66 point-like X-ray sources with unidentified
optical counterparts which were suspected AGN-candidates.

\begin{table}
\begin{center}
\caption{\label{t:grisms}
Grism properties of the EFOSC2 spectrograph. The spectral resolution 
finally achieved depends of the slit width chosen and was for the typical
value of 1.5 arcsec a 
factor of $\sim$4 higher than the values given in the table.}
{\small
\begin{tabular}[t]{ccc} \hline
Grism \# & Wavelength range (\AA ) & Dispersion (\AA /Pixel) \\ \hline
1  & 3400 -- 9200 & 8.4 \\ 
3  & 3520 -- 5470 & 1.9 \\ 
5  & 5800 -- 8400 & 2.5 \\ 
6  & 4600 -- 7200 & 2.7 \\ 
8  & 4640 -- 5950 & 1.3 \\ 
10 & 6600 -- 7820 & 1.2 \\ \hline
\end{tabular}
}
\end{center}
\end{table}

Table \ref{t:xdat} contains the X-ray positions, ROSAT-PSPC count rates and
hardness ratios of these sources.
We refer to individual sources by "RHS" (for ROSAT Hard Survey) followed by
a sequence number as given in Table \ref{t:xdat}.
As mentioned above, the RHS was a sort of pathfinder project within the 
larger ROSAT Bright Survey (RBS). The RHS-sources thus form a
subsample of the RBS, whose identification is still in progress and 
a catalogue paper is in preparation (Schwope et al.~1998). 
We, therefore, also list in Tab.~\ref{t:xdat} the sequence number of the RHS-sources
in the RBS-sample. The RBS is derived from the
ROSAT All-sky Bright Source Catalogue (1RXS, Voges et al.~1996), based on the
merged data \mbox{SASS 2} whereas at the beginning of the project, when the 
RHS was defined, only a preliminary processing step of
the RASS (\mbox{SASS 1}) was available. The advanced processing eliminated
ghosts and changed the survey positions, countrates, X-ray colors or the
information about the X-ray extent of several X-ray sources.
This notion explains why some of the RHS-sources listed in Tab.~\ref{t:xdat}
have hardness ratios smaller than the original limit of 0.5 and why other
sources have count rates smaller than the original limit of 0.2 s$^{-1}$.
Consequently, some of the sources selected for the RHS are no longer part 
of the RBS whereas new sources came in.
This means that the RHS is statistically not complete, it is, nevertheless,
a step in the identification of the whole RBS.
The positions given in Tab.~\ref{t:xdat} are the best available for the sources
in the sample, they are not exclusively drawn from the ROSAT Bright Source
Catalogue (Voges et al.~1996). Whenever possible we used coordinates from our HRI pointings.  
In order to allow a comparison between survey coordinates and those
derived from pointed observations, we include in Tab.~\ref{t:xdat} also
the 1RXS-names of the sources.
Count rates and hardness ratios are those from the 1RXS-catalog with the 
exception of RHS10, where we used the SASS1-processed survey data.
Although the existence as X-ray source of RHS10 is not in doubt it did
not pass the SASS 2 processing step and did not 
enter the 1RXS-catalog due to its large attitude uncertainty caused
by the presence of only one visible guide star.
RHS58 turned out to be an X-ray ghost (by an HRI-pointing of the authors 
and the SASS 2 process). Both sources were therefore 
eliminated from the 1RXS-catalogue.
The other sources contained in Tab.~\ref{t:xdat} without
RBS-entry (16, 21, 25, 45) are confirmed X-ray sources but 
have revised countrates below 0.2 s$^{-1}$, the defining threshold 
of the RBS sample.

\section{Optical identifications}
\label{sect_opti}
Finding charts for the likely optical counterparts of the 
X-ray emitters were produced using the APM/ROE-scans of the
POSS/ESO/SERC Schmidt-plates. 
In most cases only one or two likely
candidates for the optical counterparts 
were detected in the X-ray error circles. A few sources had
no likely optical counterpart and were proposed for short observations
with the high-resolution imager HRI onboard ROSAT in order to 
improve on the X-ray positions and to search for a possible 
X-ray extent. 

Optical observations of the AGN candidates 
took place on several occasions since 1994 using
the ESO/MPG 2.2m telescope at La Silla, the SAO 6m telescope at Zelentschuk,
the 3.5m telescope at Calar Alto and the 10m Keck telescope.
All these telescopes were equipped with low-resolution spectrographs
and CCDs as detectors, yielding  spectra with typical resolution of 
8--12\,\AA\,FWHM over the spectral range 3500 -- 9000\,\AA.
The `working horse' of our program,
the ESO/MPG 2.2m telescope, was equipped with the ESO Faint Object Spectrograph
and Camera (EFOSC2), which also allows to take CCD images of the
X-ray fields with a field of view of about $8' \times 8'$. 
For most of the  northern
RBS-sources we have only images from the digitized sky survey plates. 

With EFOSC2 at the ESO/MPG 2.2m-telescope, we took at least low resolution 
spectra with grism G\#1, which gave spectra with about 40\,\AA\,FWHM 
spectral resolution (see Table \ref{t:grisms} for a list of grism 
properties). Spectra with higher resolution using one of the other 
grisms were taken for emission line objects if telescope time was available.

During all spectroscopic observation runs spectra of standard stars were taken
which allowed a spectrophotometric calibration of the target spectra with 
estimated 50\% accuracy or better in the center of the spectra. 
Since the spectra were not in general 
taken at the parallactic angle, they suffer from light losses at the 
blue or red end of the wavelength range covered. Hence, the flux
decrease below $\sim$4500\,\AA\, seen in several spectra
should not be taken too literally.

We determined $V$-band magnitudes for our targets by
folding the spectra through a corresponding filter curve. These magnitudes are
listed in Tab.~\ref{t:optid}, which collects the results of our 
identification program. 
For extended objects like normal galaxies, cluster 
galaxies or AGNs with a strong underlying galaxy contribution even these 
rather crude brightness estimates are superior to those from the 
APM scans which gives often spurious results for these types of objects.

Since the main aim of the present study and the project in which it is embedded
is the definition of a large, statistically complete,
X-ray selected AGN sample, we did not obtain 
spectra of bright stars, $V \leq 11^m$, which are very likely coronal emitters.
This applies to the sources with RHS sequence numbers 16, 28, and 36,
respectively.

\begin{figure}[thb]
\begin{center}
\resizebox{140mm}{!}
{\includegraphics[bbllx=62pt,bblly=20pt,bburx=535pt,bbury=790pt,clip=]
{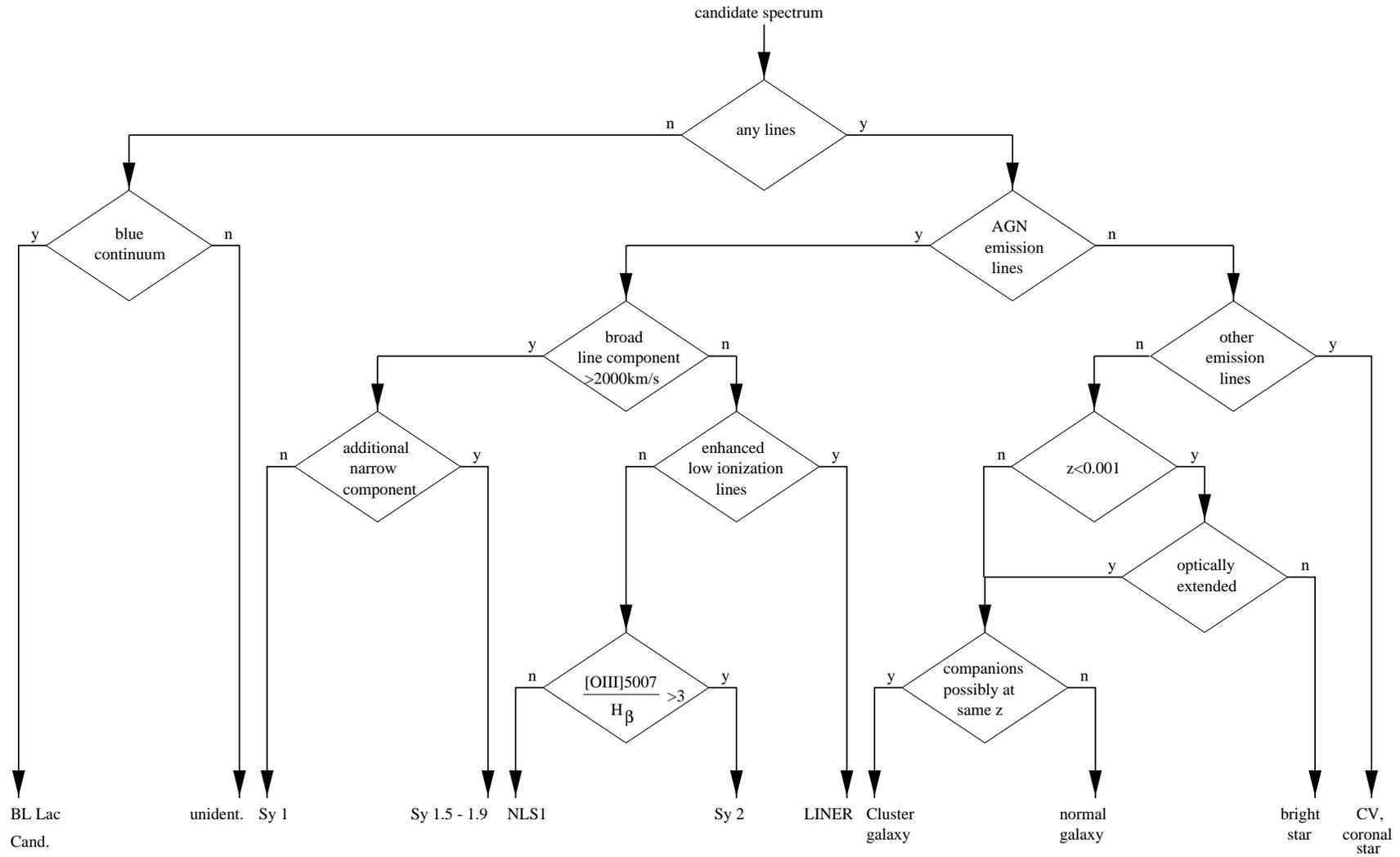}}
\end{center}
\caption{Flow chart used for classification of candidate spectra
}
\label{f:flow_chart}
\end{figure}

The optical spectra were analysed and classified in a manner similar 
to those described in related papers on identification of X-ray sources
(see e.g.~Stocke et al.~1991 for the EMSS or Appenzeller et al.~1998 for 
identification of RASS-sources in selected areas). In Fig.~\ref{f:flow_chart}
a flow chart used for classification of the objects is presented.
Additional information on the nature of the X-ray sources was derived
from X-ray data (RASS or pointed observations), which was used to 
discriminate between point-like emission from an AGN or extended
emission from a cluster.

We applied multi-component spectral fits to the emission-line AGN in order to
determine their subtype. Usually four Gaussian components were fitted 
to the H$\beta$/[O\,{\sc iii}] and the H$\alpha$/[N\,{\sc ii}] complex, thus
allowing for broad and narrow Balmer lines as well as narrow forbidden lines.
The narrow components of allowed and forbidden lines were restricted to the
same redshift and the same width. The redshifts of the broad and narrow Balmer 
lines were not determined separately but set by definition to the same values.

During the course of the project we became aware that our emission-line AGNs
are often accompanied by galaxies of similar brightness separated 
by several galaxy radii (in all cases the companions were located
outside the X-ray error circles). 
In order to test if these are physical or
apparent companions and to address the question of AGN activity
triggered by interaction we took low-resolution spectra of them if 
telescope time was available. Hence, the census of interacting Seyfert galaxies
given in the next chapter is incomplete and a more comprehensive 
study is necessary.

It is known from previous X-ray surveys (e.g.~Maccacaro et al.~1988)
that BL Lac objects are characterized by the highest $f_x/f_{\rm opt}$-ratio 
and this does apply to our sample too, 
hence, they belong optically to the faintest objects.
The discrimination between e.g.~the noisy spectrum of a normal field
galaxy, a cluster galaxy and 
a BL Lac object is therefore not easy. Indications pointing to a BL Lac 
nature of a specific source are:
absence of emission lines, optical continuum emission preferentially blue, 
contrast of the Ca-break smaller than 30\%, 
stellar-like appearance on an optical image,
point-like X-ray emission. Usually a combination of these features
applies to a BL Lac object.
The tentative classification as BL Lac is supported if a 
strong radio flux is observed and if the candidate source lies in the 
region of the $\alpha_{\rm ox} - \alpha_{ro}$ diagram usually 
populated by this 
source class. The use of such diagrams for classification was discussed by 
Stocke et al.~(1991) and Nass et al.~(1996).
A few BL Lac spectra show bumps and/or troughs in the continuum 
(associated with e.g.~the G-band or the Ca\,{\sc ii} break) or absorption 
lines from the underlying host galaxy. In these cases redshifts 
were determined by subtracting a smooth continuum and crosscorrelation 
of the residuum with a zero-velocity template spectrum.

\newpage
\clearpage

\renewcommand{\baselinestretch}{0.85}\noindent\small
\begin{table}[th]
\begin{center}
\caption{Summary of the optical identification program of the new sources in the 
RHS-sample. }
\label{t:optid}
\bigskip

\small
\begin{tabular}{|c||ccc|ccc|c|l|l|c|r|r|r|c|}
\hline
&\multicolumn{6}{c|}{Optical Position}&&&&&&&&\\
RHS &\multicolumn{3}{c|}{RA(2000)}&\multicolumn{3}{c|}{DE(2000)}&$\bigtriangleup_{ox}$&
\multicolumn{1}{c|}{Class}&
\multicolumn{1}{c|}{Type}&
\multicolumn{1}{c|}{z}&
\multicolumn{1}{c|}{$m_V$}&
\multicolumn{1}{c|}{$f_{\rm X,-12}$}&
\multicolumn{1}{c|}{$f_r$}&
\multicolumn{1}{c|}{Note}\\
No.~&$h$&$m$&$s$&$^\circ$&$'$&$''$&$''$& & & &
\multicolumn{1}{c|}{mag}&
\multicolumn{1}{c|}{cgs}&
\multicolumn{1}{c|}{mJy}&\\
\hline
 01 & 00 & 18 & 27.8 &$+29$& 47 & 32 & 02 & AGN & BL    & 0.100 & 19.4 &   4.3 &  34.7 & C,1\\
 02 & 00 & 20 & 21.7 &$+10$& 34 & 45 & 16 & AGN & Sy1   & 0.163 & 17.4 &   3.7 &       &  \\
 03 & 00 & 34 & 16.6 &$-79$& 05 & 21 & 16 & AGN & Sy1   & 0.074 & 15.4 &   5.7 &       & \\
 04 & 00 & 35 & 14.9 &$+15$& 15 & 04 & 09 & AGN & BL    &       & 16.3 &   2.1 &  18.7 & \\
 05 & 00 & 40 & 52.9 &$-07$& 42 & 10 & 10 & AGN & Sy1   & 0.055 & 16.5 &   3.7 &       & \\
 06 & 00 & 43 & 52.0 &$+24$& 24 & 21 & 18 & CLG &       & 0.083 & 17.2 &   5.8 &  49.9 & C,2 \\
 07 & 00 & 58 & 16.7 &$+17$& 23 & 14 & 15 & AGN & BL    &       & 20.3 &   3.8 &   9.2 & 1\\
 08 & 01 & 09 & 08.2 &$+18$& 16 & 07 & 04 & AGN & BL    &       & 16.6 &   4.2 &  90.8 & \\
 09 & 01 & 10 & 49.9 &$-12$& 55 & 02 & 07 & AGN & BL    & 0.234 & 17.9 &   6.4 &  17.4 & C\\
 10 & 01 & 25 & 58.0 &$+15$& 18 & 12 & 02 & AGN & Sy1   & 0.111 & 17.5 &   2.3 &       & \\
 11 & 01 & 56 & 00.2 &$+15$& 02 & 13 & 07 & AGN & BL    & 0.080 & 19.8 &   3.6 &       & C,1 \\
 12 & 02 & 06 & 39.0 &$-71$& 48 & 21 & 07 & AGN & Sy1   & 0.260 & 18.8 &   2.8 & 501.0 & 3\\
 13 & 02 & 07 & 02.2 &$+29$& 30 & 46 & 12 & AGN & Sy1.8 & 0.111 & 16.8 &   2.6 &1613.1 & 4\\
 14 & 02 & 16 & 32.1 &$+23$& 14 & 47 & 03 & AGN & BL    &       & 17.9 &   5.1 &  35.7 & 5\\
 15 & 02 & 42 & 14.6 &$+05$& 30 & 36 & 09 & AGN & Sy1   & 0.069 & 16.0 &   5.1 &   3.1 & I\\
 16 & 02 & 45 & 24.1 &$+14$& 35 & 22 & 05 & STAR&       &       &  7.9 &   2.3 &       & 6\\
 17 & 03 & 00 & 08.0 &$+16$& 30 & 15 & 10 & AGN & Sy1   & 0.035 & 16.4 &   5.0 &       & \\
 18 & 03 & 01 & 38.2 &$+01$& 55 & 15 & 05 & CLG &       & 0.170 & 18.0 &   2.9 & 393.6 & C,2\\
 19 & 03 & 03 & 30.1 &$+05$& 54 & 17 & 08 & AGN & BL    & 0.196 & 17.9 &   3.9 &  29.6 & C\\
 20 & 03 & 14 & 23.9 &$+06$& 19 & 57 & 03 & AGN & BL    &       & 17.9 &   8.7 &  29.3 & \\
 21 & 03 & 16 & 12.8 &$+09$& 04 & 43 & 04 & AGN & BL    &       & 18.2 &   2.3 &  55.4 & \\
 22 & 03 & 27 & 39.2 &$-58$& 09 & 50 & 13 & STAR& dKe   &       & 11.7 &   2.5 &       & \\
 23 & 03 & 34 & 24.4 &$-15$& 13 & 40 & 14 & AGN & Sy1.5 & 0.035 & 15.8 &   3.0 &   4.7 & I,7\\
 24 & 03 & 52 & 57.4 &$-68$& 31 & 19 & 02 & CLG:&       & 0.087 & 17.1 &   7.9 & 175.0 & C,8\\
 25 & 04 & 41 & 53.9 &$-08$& 26 & 34 & 10 & AGN & Sy1:  & 0.044 & 15.7 &   2.5 &       & 9\\
 26 & 05 & 35 & 26.7 &$-43$& 22 & 45 & 09 & AGN & Sy1.5 & 0.065 & 16.8 &   1.9 &       & \\
 27 & 05 & 43 & 57.3 &$-55$& 32 & 08 & 02 & AGN & BL    &       & 17.4 &   8.3 &       & \\
 28 & 08 & 27 & 13.9 &$+41$& 28 & 38 & 08 & STAR&       &       & 10.6 &   2.3 &       & 6\\
 29 & 08 & 38 & 11.1 &$+24$& 53 & 44 & 08 & AGN & Sy1:  & 0.028 & 16.3 &   2.4 &  65.6 & I,10\\
 30 & 09 & 47 & 12.5 &$+76$& 23 & 14 & 04 & AGN & LINER & 0.354 & 19.2 &   3.0 &  22.4 & \\
 31 & 10 & 56 & 06.6 &$+02$& 52 & 13 & 06 & AGN & BL    & 0.235 & 19.5 &   6.8 &   4.3 & \\
 32 & 11 & 24 & 07.3 &$+06$& 12 & 47 & 13 & AGN & Sy1.5-1.8&0.036&16.2 &   4.3 &       & \\
 33 & 11 & 45 & 16.3 &$+79$& 40 & 52 & 03 & AGN & Sy1/NLS1&0.006& 15.2 &   4.5 &       & \\
 34 & 12 & 07 & 11.5 &$-17$& 46 & 06 & 04 & AGN & BL:   &       & 20.5 &   3.0 &   4.3 & 11\\
 35 & 12 & 13 & 23.0 &$-26$& 18 & 07 & 06 & AGN & BL    & 0.278 & 19.0 &   3.9 &   7.3 & \\
 36 & 12 & 48 & 43.1 &$+06$& 54 & 22 & 03 & STAR& G0    &       &  8.9 &   3.7 &       & 12\\
 37 & 13 & 31 & 13.7 &$-25$& 24 & 02 & 04 & AGN & Sy1.5-1.8&0.026&15.7 &   7.7 &  12.9 & 13\\
 38 & 13 & 54 & 20.0 &$+32$& 55 & 50 & 03 & AGN & Sy1   & 0.026 & 15.4 &   7.0 &   3.7 & 14\\
 39 & 14 & 19 & 22.2 &$-26$& 38 & 41 & 06 & AGN & Sy1   & 0.022 & 14.9 &  12.1 &  13.4 & I,15\\
 40 & 14 & 37 & 03.4 &$+23$& 42 & 27 & 10 & CV  & NL    &       & 20.5 &   2.6 &       & 16\\
 41 & 14 & 45 & 06.2 &$-03$& 26 & 12 & 03 & AGN & BL    &       & 17.2 &   2.9 &  21.6 & \\
 42 & 14 & 47 & 26.0 &$+08$& 27 & 12 & 22 & CLG:&       & 0.190 & 19.9 &   2.6 &       & 17\\
 43 & 14 & 47 & 38.5 &$-08$& 33 & 26 & 03 & AGN & Sy1   & 0.196 & 16.4 &   3.2 &       & \\
 44 & 14 & 49 & 32.7 &$+27$& 46 & 21 & 02 & AGN & BL    & 0.225 & 18.6 &   3.9 &  90.1 & \\
 45 & 15 & 06 & 37.0 &$-05$& 40 & 05 & 11 &     &       &       & 20.0 &   1.5 &       & \\
 46 & 15 & 08 & 42.8 &$+27$& 09 & 10 & 04 & AGN & BL    &       & 18.5 &   3.3 &  39.9 & \\
 47 & 15 & 16 & 18.5 &$-15$& 23 & 45 & 04 & AGN & BL    &       & 19.6 &   5.7 &   8.0 & \\
 48 & 15 & 23 & 46.0 &$-00$& 44 & 25 & 02 & STAR& dKe   &       & 11.3 &   4.3 &       & \\
 49 & 15 & 24 & 51.4 &$-10$& 05 & 59 & 07 & AGN & Sy1   & 0.144 & 17.3 &   4.3 &       & 18\\
 50 & 15 & 30 & 49.5 &$+20$& 26 & 05 & 04 & AGN & Sy1/NLS1&0.216& 16.7 &   2.6 &       & \\
 51 & 15 & 31 & 41.2 &$+20$& 19 & 29 & 06 & AGN & Sy1/NLS1&0.051& 17.2 &   2.5 &       & \\
 52 & 15 & 32 & 53.7 &$+30$& 20 & 55 & 09 & AGN & LINER & 0.362 & 19.1 &   2.6 &  23.8 & \\
 53 & 15 & 40 & 15.7 &$+81$& 55 & 05 & 01 & AGN & BL    &       & 17.6 &   4.5 &  69.9 & 19\\
 54 & 17 & 02 & 44.1 &$+72$& 53 & 28 & 07 & AGN & Sy1   & 0.053 & 16.4 &   4.6 &   4.3 & 20\\
 55 & 17 & 17 & 07.0 &$+29$& 31 & 20 & 01 & CLG &       & 0.275 & 19.6 &   3.1 &   3.2 & 21\\
 56 & 20 & 39 & 27.0 &$-30$& 18 & 53 & 03 & AGN & Sy1/NLS1&0.080& 16.0 &   4.8 &   5.9 & \\
 57 & 21 & 50 & 15.5 &$-14$& 10 & 50 & 03 & AGN & BL    & 0.229 & 18.5 &   6.6 &  68.3 & C\\
 60 & 22 & 43 & 41.5 &$-12$& 31 & 38 & 33 & AGN & BL    & 0.226 & 20.0 &   3.1 &  10.5 & C\\
 61 & 23 & 25 & 54.4 &$+21$& 53 & 16 & 07 & AGN & Sy1   & 0.120 & 15.9 &   6.2 &   6.0 & I\\
 62 & 23 & 41 & 06.3 &$+09$& 38 & 09 & 05 & AGN & Sy1   & 0.041 & 16.5 &   2.7 &   5.1 & I\\
 63 & 23 & 43 & 13.5 &$-36$& 37 & 54 & 04 & AGN & Sy1   & 0.622 & 16.0 &   0.6 &  10.3 & \\
 64 & 23 & 47 & 28.7 &$+24$& 27 & 46 & 03 & AGN & Sy1   & 0.157 & 17.1 &   3.1 &       & I\\
 65 & 23 & 50 & 17.9 &$-05$& 59 & 28 & 01 & AGN & BL    & 0.515 & 19.5 &   3.2 &  25.1 & C\\
 66 & 23 & 55 & 48.2 &$+25$& 30 & 32 & 05 & AGN & Sy1   & 0.057 & 16.1 &   2.8 &  12.4 & I,22\\
\hline
\end{tabular}
\end{center}

\normalsize

\end{table}
\clearpage
\rfnote{\bf Notes:}

\rfnote{I: Interacting Seyfert galaxy}
\rfnote{C: Redshift determined by crosscorrelation with a zero-velocity 
template spectrum}
\rfnote{1: Nass et al.~(1996)}
\rfnote{2: Crawford et al.~(1995)}
\rfnote{3: PKS0205-720}
\rfnote{4: 3C59}
\rfnote{5: Object 'A' on the chart with BL Lac-type spectrum, object 'B' has
	faint, red spectrum without emission lines, probably a 
	normal galaxy}
\rfnote{6: Identification inferred from the Digitized Sky Survey}
\rfnote{7: Very active companion galaxy 'B' with Seyfert-type spectrum,
	companion 'C' has Seyfert-type spectrum too, but at a redshift
	of 0.15, and does not belong to the group}
\rfnote{8: Galaxies 'A' and 'B'' have redshifts of 0.087 and 0.088, respectively,
	and are probable cluster galaxies. No X-ray extent in the RASS, 
	rather small 
	Ca\,{\sc ii} break with contrast $\sim$30\%, therefore BL Lac 
	classification not completely excluded}
\rfnote{9: 1ES0439-085}
\rfnote{10: MRK1218, Keel (1996)}
\rfnote{11: Uncalibrated, noisy spectrum without emission lines; single 
	faint optical counterpart only}
\rfnote{12: SIMBAD}
\rfnote{13: ESO509-G038}
\rfnote{14: MRK663}
\rfnote{15: ESO511-G030}
\rfnote{16: H Balmer and He\,{\sc i} emission, no He\,{\sc ii}}
\rfnote{17: Spectrum of a normal galaxy, several faint galaxies around,
	no X-ray extent, the brighter objects below the X-ray error 
	circle are normal stars with no indication for activity. The 
	spectrum of the galaxy has a rather small contrast of $\sim$30\%
	in  the 	Ca\,{\sc ii} break, we cannot exclude therefore
	a BL Lac classification}
\rfnote{18: HE1522-0955}
\rfnote{19: 1ES1544+820}
\rfnote{20: UGC10697}
\rfnote{21: Point-like X-ray emission in RASS, extended in HRI-pointing}
\rfnote{22: VV 697, IRAS\,23532+2513, Zou et al.~(1995)}\\[3ex]
\renewcommand{\baselinestretch}{1.0}\normalsize
\noindent

\section{Results}
The results of our spectroscopic observations are summarized in 
Tab.~\ref{t:optid}, in the appendix we show the spectra on which our 
classification is based (Figs.~6 and 7).
We do not show the spectra of the galaxy clusters and of the active 
stars. Finding charts are reproduced in Fig.~5.
In total we obtained spectra for 60 of the 
original 66 fields which formed the RHS. We did not obtain spectra
for the bright stars, which are the likely counterparts of RHS16, RHS28,
and RHS36, but we regard their identification as coronal emitters as certain.
After the start of the optical identification work we became aware, that a few
fields we were working on did not reveal likely counterparts. We then 
observed those fields with the high resolution imager onboard ROSAT
in order to improve on the accuracy of the X-ray positions and to
search for extended X-ray emission. As a result 
of these short HRI-pointings (typical exposure time 2\,ksec) 
the positions of some X-ray sources were shifted considerably. One of the
original sources turned out to be a ghost image (RHS58), the revised positions
of two further sources were matching those of catalogued objects
(RHS36, a bright star, and RHS59, a Seyfert 1.8 galaxy) and thus were 
not observed  spectroscopically.

\begin{figure}[hbt]
\resizebox{115mm}{!}
{\includegraphics[angle=-90]
{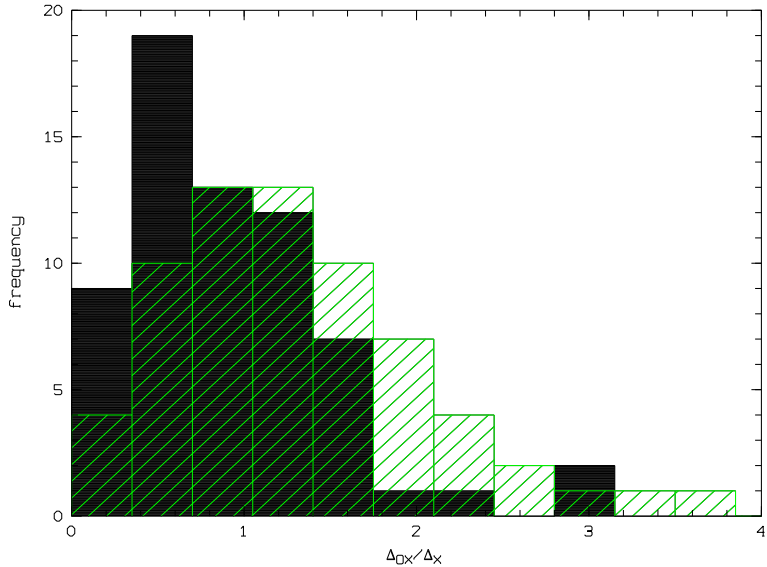}}
\hfill
\parbox[t]{55mm}{
\caption{
Normalized separations between optical and X-ray positions in units 
of the 1-$\sigma$ positional uncertainty of the X-ray sources (black 
histogram). The dashed histogram represents a Gaussian offset distribution.
}
\label{f:diff_pos}}
\end{figure}

After all, only the field of RHS45 remained unobserved.
Since the SASS-2 processed RASS gives a countrate for that object 
below 0.2\,s$^{-1}$
we finally did not attempt to classify this
object spectroscopically.
Some other identifications remain ambiguous due to low signal-to-noise 
in our spectra.
This applies exclusively to objects without emission 
lines, hence, the identification of or discrimination between BL Lac objects, 
galaxy clusters and normal galaxies might in some cases be doubtful. These
cases are discussed in the notes to Tab.~\ref{t:optid}. 

Table~\ref{t:optid} is organized in the following way: the first column 
is the RHS-sequence number, columns 2 to 7 contain the equatorial coordinates
of the objects that have been spectroscopically observed, column 8 shows the
positional difference between the optical and the X-ray coordinates as listed 
in Tab.~\ref{t:xdat}, columns 9 and 10 labeled 'Class' and 'Type' give the 
identification of the X-ray sources, column  11 (labeled $z$) holds the redshift
as far as it could be determined, column 12 (labeled $m_V$) lists the 
optical brightness, column 13 the integrated X-ray flux between 0.5 
and 2.0\,keV, column 14 the radio flux at 1.4 GHz and the last column 
labeled 'Note' references notes 
at the end of the table. Most optical brightness values are derived
from actually observed spectra, for RHS34 and the bright stars magnitudes 
from the APM-scans are listed. The X-ray flux given in the table 
is the unabsorbed flux in units of
$10^{-12}$ erg $^{-1}$ cm$^{-2}$ s$^{-1}$ and 
has been computed assuming a power law spectrum with energy index $-1$
(Hasinger et al.~1993) taking into account the galactic column 
density $N_{\rm H}$ towards the source.
The radio fluxes are taken from the NRAO/VLA Sky Survey (NVSS, 
Condon et al.~1998), except those for RHS12 and \#24 which were 
taken from the Parkes-MIT-NRAO survey (PMN, Griffith \& Wright 1993).
The PMN-fluxes, given at 4.85 GHz, were transformed to those at 1.4 GHz
assuming a power law with energy index $-0.7$ (Stocke et al.~1991).

The distribution of the separations between the X-ray sources
and the identified optical counterparts, normalized to 1$\sigma$ positional 
uncertainty, 
is shown in Fig.~\ref{f:diff_pos}.
The result expected for a 2d-Gaussian distribution of positional offsets 
is included in the figure.
The 90\%-confidence error radius is derived from the 1-$\sigma$ positional
uncertainties $\Delta {\rm RA}$ and $\Delta {\rm DEC}$ given in the 1RXS
catalogue as $r_{90} = 1.65 \sqrt((\Delta {\rm RA}))^2 + (\Delta {\rm DEC})^2
\simeq 1.65 \sigma $ ($\sigma$ as given in Tab.~\ref{t:xdat}). 
There is no excess of the observed distribution over the Gaussian, the 
contrary seems to be true. This might be slightly indicative of too large 
estimated errors for the X-ray positions. In the diagram the observed
cluster galaxies have been included. Since the center of the X-ray emission 
of a cluster and the optical position of the observed galaxy
do not necessarily coincide, the good positional coincidence 
as expressed in Fig.~\ref{f:diff_pos} is even more surprising. Only 
one cluster, RHS06, is located at $\Delta_{\rm OX} / \Delta_{\rm X} = 3$.
The statistics of the normalized separations between X-ray and optical 
positions confirms that the proposed counterparts are very likely the 
correct identifications.

In sum, only one X-ray source, which after SASS-2 processing of the RASS 
does not further obey  the original selection criteria, remains unidentified.
All others have almost secure optical identifications, a few cases remain
ambiguous (see notes to Tab.~\ref{t:optid}). Our sample contains 5 coronally
active stars and one faint cataclysmic variable star, all other 59 X-ray sources 
are of extragalactic origin. Not so surprisingly since we tried to exclude 
this type of X-ray source we find only five (\#6, \#18, \#24, \#42, \#55)
clusters of galaxies in our sample.
After SASS-2 processing RHS6 has migrated into the 
upper right regime of Fig.~\ref{f:ext}, RHS55 lies on and RHS42 near to 
the dotted dividing line between extended and point-like objects whereas 
RHS18 and RHS24 lie well inside the regime usually populated by AGNs. An 
HRI-pointing, however, has revealed clearly extended X-ray emission from 
RHS18 and the identification as cluster can be regarded as certain. 
Short pointed X-ray observations of RHS24 and RHS42 are necessary in order to 
ascertain their identification. For the time being, an identification as
BL Lac object cannot be excluded for these two sources.

\begin{figure}[thb]
\resizebox{115mm}{!}
{\includegraphics[bbllx=52pt,bblly=82pt,bburx=523pt,bbury=565pt,angle=-90,clip=]
{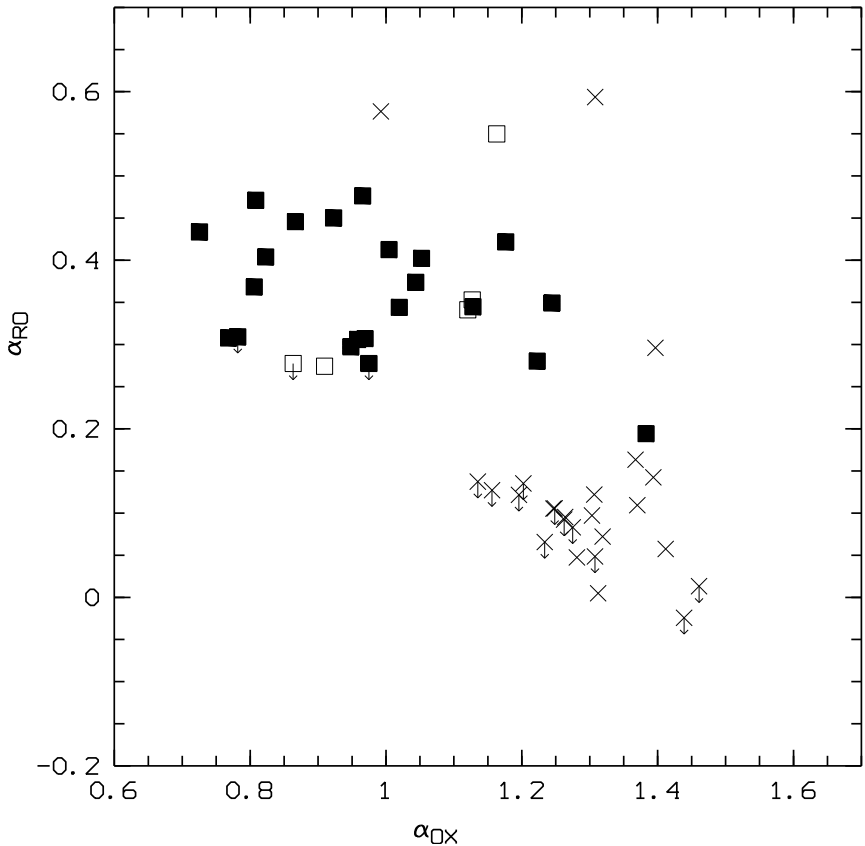}}
\hfill
\parbox[t]{55mm}{
\caption{
The $\alpha_{\rm OX} - \alpha{\rm RO}$ diagram of the extragalactic
sources identified in this paper. Crosses indicate emission-line
AGNs, open squares indicate clusters, filled squares BL Lac objects. 
Arrows denote objects where only upper limits for the radio flux
are known. 
}
\label{f:aro_aox}}
\end{figure}

The large majority of the RHS X-ray sources turned out to be AGNs 
(Seyfert galaxies and BL Lac objects). 
The fraction of BL Lac objects, about one third of the total number of RHS-sources,
is high. These sources have not been noticed  previously
due to their low optical brightness and they pop up here 
because of their high energy output in the X-ray regime. The use 
of the high $f_X/f_{\rm opt}$-ratio in order to search new BL Lac objects
in X-ray surveys has been discussed by Nass et al.~(1996). 
In Fig.~\ref{f:aro_aox} we show the locations of our new identifications
in the $\alpha_{\rm ox} - \alpha_{\rm ro}$ plane. For calculation of the 
corresponding values of our sources we were using the radio flux at 1.4\,GHz,
the optical flux at 6000\,\AA, and the X-ray flux at 1\,keV.
The objects labeled in 
Tab.~\ref{t:optid} as BL Lac objects are found in a narrow strip centered
on $\alpha_{\rm ro} \simeq 0.38$ which supports our identification
(cf.~Nass et al.~1996, Stocke et al.~1991). The only exception is RHS04
which is found in the vicinity of the radio quiet emission-line AGNs.
From its optical spectrum it is, however, a certain BL Lac object.
The diagram gives no further clue to the nature of the RHS24 and 42, 
which are both found in the region populated by BL Lac objects,
since the regions populated by clusters and BL Lacs have a large 
overlap.

Most of the AGNs in the RHS sample are emission line objects of different 
subclasses, including two LINERs, 5 Seyfert galaxies of types $1.5-1.8$, 
and 4 possible narrow-line Seyfert 1 galaxies (NLS1) 
but the majority of them being of Seyfert type 1.
Some of the objects listed in Tab.~\ref{t:optid} were previously catalogued,
we mention e.g.~Mrk1218, Mrk663 or VV697 (RHS29, 38, and 66, 
respectively). However, when the project was started, their nature as
Seyfert galaxies or their redshifts were unknown and these galaxies were 
included therefore in our identification program. With the exception
of RHS12 and 13 all emission line AGNs are radio quiet.

Among the new Seyfert-type AGNs a surprisingly large fraction is in 
interaction with a nearby single companion or in a small group. We have 
eight confirmed examples (marked 'I' in column 'Note' of Tab.~\ref{t:optid})
with a redshift difference between Seyfert and companion $\Delta z < 0.001$ 
and a few more candidates. CCD images and spectra of the confirmed cases
are shown in Fig.~8. From the APM finding
charts or from CCD images
we tentatively identified a couple of further possible cases (\#12, \#13, 
\#17, \#25, \#37, \#38 and \#43). It turned out by low-resolution spectroscopy 
of the 
galaxies labeled 'B' on the finding charts (Fig.~5) of
RHS12 and 43, respectively, that these galaxies are apparent 
companions only (RHS12B: normal galaxy, z=0.049; RHS43B: emission line
galaxy at z=0.129). 
Some of the companion galaxies itself show indication of activity
by e.g.~enhanced star formation visible in their optical spectra 
as prominent H$\alpha$ emission lines (RHS 15, 23, 61, 66).
The eight certain cases together with the five possible cases 
(RHS13, 17, 25, and 38) represent between 26\%{} and 42\%{} of all new 
emission-line AGNs. This large fraction of interacting Seyfert galaxies
is clearly much higher than that found in an optically selected sample
(Rafanelli et al.~1995 found a 12\% excess of galaxy-Seyfert pairs over 
galaxy-galaxy pairs) and supports recent scenarios of tidal 
interaction triggering starburst and Seyfert activity (see e.g.~Cavaliere \&
Vittorini 1998, Bahcall et al.~1997).

\acknowledgements{We thank Klaus Reinsch (G\"ottingen) for providing
the spectra for RHS33 and RHS54. We acknowledge many helpful comments
of the referee T.~Maccacaro.
The ROSAT project is supported by the Bundesministerium f\"{u}r 
Bildung, Wissenschaft, Forschung und Technologie (BMBF/DLR) and the 
Max-Planck-Gesellschaft. We thank the ROSAT team for performing the All-Sky
Survey and producing the RASS Bright Source Catalogue. 
This research has made use of the SIMBAD database operated at
CDS, Strasbourg, France, and the NASA/IPAC Extragalactic database (NED)
operated by the Jet Propulsion Laboratory, California Institute of 
Technology under contract with the National Aeronautics and Space 
Administration. Identification of the RASS X-ray sources was greatly
facilitated by use of the finding charts based upon the COSMOS
scans of the ESO/SERC J plates performed at the Royal Observatory 
Edinburgh and APM catalogue based on scans of the red and blue POSS plates
performed at the Institute of Astronomy, Cambridge, UK. 

Based in part 
on photographic data of the National Geographic Society -- Palomar
Observatory Sky Survey (NGS-POSS) obtained using the Oschin Telescope on
Palomar Mountain.  The NGS-POSS was funded by a grant from the National
Geographic Society to the California Institute of Technology.  The
plates were processed into the present compressed digital form with
their permission.  The Digitized Sky Survey was produced at the Space
Telescope Science Institute under US Government grant NAG W-2166.

This work has been supported in part by the DLR (former DARA GmbH)
under grant 50 OR 9403 5.
}



\addresses
\rf{Hermann Brunner, Astrophysikalisches Institut Potsdam, 
An der Sternwarte 16, D-14482 Potsdam, Germany, \\e-mail: HBrunner@aip.de}
\rf{Jens-Uwe Fischer, Astrophysikalisches Institut Potsdam, 
An der Sternwarte 16, D-14482 Potsdam, Germany, \\e-mail: JUFischer@aip.de}
\rf{G\"unther Hasinger, Astrophysikalisches Institut Potsdam, 
An der Sternwarte 16, D-14482 Potsdam, Germany, \\e-mail: GHasinger@aip.de}
\rf{Axel Schwope, Astrophysikalisches Institut Potsdam, 
An der Sternwarte 16, D-14482 Potsdam, Germany, \\e-mail: ASchwope@aip.de}
\rf{Thomas Boller, Max-Planck-Institut f\"ur extraterristrische Physik, 
D-85740 Garching, Germany, \\ e-mail: bol@mpe-garching.mpg.de}
\rf{Joachim Tr\"umper, Max-Planck-Institut f\"ur extraterristrische Physik, 
D-85740 Garching, Germany, \\e-mail: jtrumper@mpe-garching.mpg.de}
\rf{Wolfgang Voges, Max-Planck-Institut f\"ur extraterristrische Physik, 
D-85740 Garching, Germany, \\e-mail: whv@mpe-garching.mpg.de}
\rf{Sergej Neizvestny, SAO RAS, Nizhnij Arkhyz, Zelenchukskaya,
Karachaevo-Cherkesia, Russia, 357147}
%

\clearpage
\end{document}